# Time pressure and honesty in a deception game

Valerio Capraro[1], Jonathan Schulz[2], David G. Rand[3]

[1]Middlesex University London, [2]Harvard University, [3]Yale University

**Abstract**

Previous experiments have found mixed results on whether honesty is intuitive or requires deliberation. Here we add to this literature by building on prior work of Capraro (2017). We report a large study (N=1,389) manipulating time pressure vs time delay in a deception game. We find that, in this setting, people are more honest under time pressure, and that this result is not driven by confounds present in earlier work.

**Introduction**

Whether honesty is intuitive or requires deliberation is a current topic of debate. Two early studies using time constraints to manipulate cognitive mode found that honesty requires deliberation (Gunia et al, 2012; Shalvi et al, 2012). This conclusion was challenged by three more recent studies, two of which found the opposite effect, that time pressure promotes honesty (Capraro, 2017; Lohse et al, 2018), while the third one found a null effect (Barcelo & Capraro, 2017). Studies using different cognitive processing manipulations, such as conceptual priming of intuition (Cappelen et al, 2013), ego-depletion (Gino, Schweitzer, Mead & Ariely, 2011), and cognitive load (van't Veer, Stel & van Beest, 2014) have also led to mixed results.

Here, we add further evidence on this issue. We build on the prior work of Capraro (2017), who used an experimental lying paradigm similar to Biziou-van-Pol et al (2015). In Capraro (2017), participants are randomly assigned to a group (either Group 1 or Group 2). They have to decide between two options: either honestly telling the group number they are assigned to or dishonestly telling the number of the other group. If they report the true number of the group they are assigned to, then both the decision maker and a randomly matched participant will get $0.10; otherwise the decision maker will get $0.20 and the matched participant will get $0.09. Half of the participants are asked to decide within 5 seconds (time pressure); while the other half are asked to think carefully for at least 30 seconds before deciding (time delay). To avoid that participants decide which group number to report before this experimental manipulation, they are not initially told the number of the group they are assigned to. This piece of information is provided directly in the decision screen. Since the number of the group is not initially provided, the payoff maximizing strategy – not reporting the true number of their group – is not

immediately accessible. Capraro (2017) found that, in this context, time pressure promotes truth-telling and used this finding to support the hypothesis that honesty is intuitive.

However, a closer look at the design reveals that it contains two confounds. First, in Capraro (2017) dishonesty increases social welfare: An honest participant obtained $0.10 and $0.10 was allocated to the other, matched participant ($0.20 total), while a dishonest participant obtained $0.20 and $0.09 was allocated to the paired participant ($0.29 total). Here, we adjust incentives such that the dishonest choice does not increase social welfare, i.e., the same total amount was distributed independently of whether or not the participant made a dishonest choice or not.

The second, and, arguably, more critical confound is that in Capraro (2017) participants are always shown the number of their own group. This could in principle invalidate the conclusions of Capraro (2017): it is possible that time pressure merely makes people more likely to report whatever number they are presented with, because that number is more immediately accessible. This could explain the pattern of results in the prior experiment even if honesty was not actually intuitive.

Here we test this alternative explanation by employing a modified design in which we vary whether participants are shown the number of their own group or the number of the group of the matched participant. We conduct a 2x2 experiment where, along with a cognitive mode manipulation (*Time Pressure* vs *Time Delay*), we also vary the number presented (*Own Group Shown* vs *Other Group Shown*). If honesty is really intuitive, then time pressure should increase honesty in both the O*wn Group Shown* and *Other group shown* condition. If, alternatively, it is just that time pressure makes people write down whatever number is presented to them, then there will be an interaction between time pressure and group shown, such that people under time pressure are more honest in the "own group shown" condition and more dishonest in the "other group shown" condition. To summarize, we pit the following two hypotheses against each other:

*Intuitive Honesty Hypothesis.* Time pressure increases honesty regardless of whether participants are shown their own or the matched participant's group number.

*Alternative Hypothesis.* Time pressure increases honesty in the "own group shown" condition, but increases dishonesty in the "other group shown" condition.

**Experimental design and procedure**

*Protocol*

We collected data in two sessions.[1] All participants initially read the same set of instructions, in which they were informed that they had been paired with another anonymous participant, that there were two groups, Group 1 and Group 2, and that they will be randomly assigned to either of these groups, whereas the matched participant will be assigned to the other group. Participants were also informed that their job was to report the number of the group they were assigned to and that payoffs are determined as follows: if they report the number of the group they are assigned to, then they earn $0.10 and the matched participant earns $0.10; if they report the number of the other group, then they earn $0.15 and the matched participant earns $0.05. After these general instructions, participants were asked two comprehension questions, one regarding the choice that maximizes their payoff, and one regarding the choice that maximizes the matched participant's payoff. Participants failing either or both comprehension questions were automatically excluded from the survey. Participants who passed this attention check were randomly divided in 4 conditions (2x2 design). In the *Time Pressure/Own Group Shown* condition, participants were told their group number and asked to report the group they were assigned to within 5 seconds. A timer was shown to pressure the participants. However, participants could take as long as they wanted to decide. Responses were collected using a blank box. In the condition *Time Pressure/Other Group Shown* participants were told the matched participant's group number and asked to report their own group number within 5 seconds. Finally, the conditions *Time Delay/Own Group Shown* and *Time Delay/Other Group Shown* were similar to the two "pressure" conditions with the difference that this time participants were asked to think carefully for at least 30 seconds before making a choice. The button to submit the choice appeared after 30 seconds, so participants were not allowed to submit their choice before.[2] Thus, in all "*Own Group Shown*" conditions the dishonest but payoff maximizing choice was to report a number different than the one displayed, while in all "*Other Group Shown*" conditions the payoff maximizing choice was to report the number displayed.[3]

---

[1] The second session was similar to the first. Both sessions contained all treatments, participants could only participate in one session, and roughly the same number of participants participated in each session. The difference between the two sessions is that, in the second session, our experiment was appended to another experiment. This was done to save research funds. In the appendix (Table A.1) we test for session differences and find no evidence that this impacts our findings.

After making their decision, participants entered a standard demographic questionnaire, at the end of which they were provided with a completion code, with which they could claim their payment. After the experiment ended, bonuses were computed and paid on top of the participation fee ($0.50). The matched participants were selected at random from the same sample. That is, each participant received a bonus both from their choice as decision maker, and from being in the role of the "other participant" for a different participant (when making their decisions, participants were not informed that they would also be in the "other participant" role to avoid having this affect their choice). Verbatim instructions are reported in the Appendix.

*Participants*

Participants, living in the US at the time of the experiments, were recruited using the online labor market Amazon Mechanical Turk (AMT). AMT experiments are easy and cheap to implement, because participants take part from their homes in an online incentivized survey that takes no more than a few minutes. This allows experimenters to decrease the stakes at hand without compromising the quality of the results. Numerous studies have indeed shown that data gathered using AMT are of no less quality than data collected on the standard physical lab (Arechar, Molleman & Gächter, 2018; Brañas-Garza, Capraro & Rascón-Ramírez, 2018; Goodman, Cryder & Cheema, 2013; Horton, Rand & Zeckhauser, 2011; Paolacci, Chandler & Ipeirotis, 2010). To assure data quality, we restricted the survey to M-Turkers with a minimum approval rate in prior experiments of 90%[4]. In addition, we asked comprehension questions. A total of 1,389 M-turkers (mean age = 36.3, female = 48.1%) out of 2,033 passed the comprehension questions and participated in our experiment (653 in the first session and 644 in the second session). Another positive feature of the M-Turk participants relative to undergraduate student samples is that they vary in their educational attainment. In our case: 13.8% had obtained a graduate degree, 42.3% a bachelor's degree, 26.8% attended college, 4.8% had vocational training, while 12.4% had a high-school degree or less.

---

[2] The average response time of participants deciding under time pressure was significantly lower than the average response time of participants acting under time delay (10.52s vs 41.40s, rank-sum, $p < .001$), which shows that our manipulation was successful, although only 8.4% of subjects under time pressure managed to respond within five seconds.
[3] Within each condition we randomly counter-balanced the group numbers, i.e., 50% of the time the own group was group number 2, while 50% of the time the own group was group number 1. This avoids confounding due to some sort of "attachment" to numbers.

**Results**

Linear regression[5] predicting *Honesty* as a function of *Time Pressure* (1 if the participant was in the time pressure conditions, 0 otherwise) and *Other Group Shown* (1 if the participant was shown the group of the matched participant, 0 if the participant was shown their own group) find that both these variables have a significant positive effect on honesty (Table 1, Column (1)). These positive effects are robust after controlling for sex, age, and education (Table 1, Column (2)). Interestingly, adding these demographic variables reveals a significant effect of gender, such that females tend to be more honest than males. In sum, participants under time pressure are more honest than those acting under time delay, and participants who are shown the group of the matched participant are more honest than those who are shown their own group.

Next, we test our main research question. In Column (3) and Column (4) we repeat the previous linear regressions by adding the interaction term *Time Pressure X Other Group Shown*, without and with control on demographic characteristics, respectively. Note that the *Alternative Hypothesis* predicts that the interaction term is significant and negative (and sufficiently large to yield a significant negative net effect of time pressure on the *Other Group Shown* condition), while the *Intuitive Honesty Hypothesis* predicts that the interaction term is not significant and close to zero. As can be seen in Table 1, the interaction term is not significant (p=0.241, Column (3)), but also not so close to zero. If we split the sample by whether participants were shown their own group or the group of the matched participant, we find positive coefficients of time pressure on honesty in both conditions, albeit only statistically significant in the *Own Group Shown* condition (Table 1, Columns (5-6)) and not in the *Other Group Shown* condition (Table 1, Columns (7-8)). Specifically, in the *Own Group Shown* condition, honesty under time pressure is significantly higher than honesty under time delay (65.5% vs 56.8%, p = 0.017, Column 5), while in the *Other Group Shown* condition honesty under time pressure is numerically but not statistically higher than honesty under time delay (72.5% vs 69.7%, p = 0.415, Column 7).[6]

---

[4] At the end of an AMT experiment, the experimenter can either approve or reject a participant. Rejection generally happens when a participant tries to submit the survey using a fake completion code. This could happen, for example, because the participant has failed the comprehension questions and did not receive a completion code. However, AMT allows experimenters to decide the minimum approval rate of participants to be recruited.

[5] We report linear regression models despite the binary outcome variable because estimates for interaction effects are hard to interpret in logit and probit models (Ai and Norton, 2003; Greene, 2010). None of the predicted values of the linear probability models fall outside the [0,1] interval. Nevertheless, we report the outcomes of logit regression in the appendix (Table A.3).

[6] In the appendix we test for differences between the two sessions and found no evidence for it (Table A.1). There, we also report on regression analyses where instead of sample split, we interacted all variables with the *Other Group Shown* condition. This allows us to investigate whether the demographic characteristics like age or gender interact with the *Other Group Shown* condition (Table A.2, Column (1)). Furthermore, we investigated whether gender interacts with our two conditions (Table A.2. Columns (2) and (3)). We do not find evidence that the demographics interact with the four conditions.

Thus, we did not find support for the key predictions of the *Alternative Hypothesis:* the negative interaction is not statistically significant, and the coefficient of time pressure in the *Other Group Shown* condition is positive rather than negative (although not significant).

|  | Honesty | | | | | | | |
|---|---|---|---|---|---|---|---|---|
|  | Full sample | | | | Own Group Shown sub-sample | | Other Group Shown sub-sample | |
|  | (1) | (2) | (3) | (4) | (5) | (6) | (7) | (8) |
| Time Pressure | 0.058** | 0.056** | 0.087** | 0.089** | 0.087** | 0.091** | 0.028 | 0.029 |
|  | (0.025) | (0.025) | (0.037) | (0.036) | (0.037) | (0.036) | (0.035) | (0.035) |
| Other Group Shown | 0.099*** | 0.097*** | 0.129*** | 0.130*** |  |  |  |  |
|  | (0.025) | (0.025) | (0.036) | (0.036) |  |  |  |  |
| Time Pressure X Other Group Shown |  |  | -0.059 | -0.066 |  |  |  |  |
|  |  |  | (0.050) | (0.050) |  |  |  |  |
| Female |  | 0.057** |  | 0.057** |  | 0.078** |  | 0.035 |
|  |  | (0.025) |  | (0.025) |  | (0.037) |  | (0.035) |
| Age |  | 0.002** |  | 0.002** |  | 0.003** |  | 0.002 |
|  |  | (0.001) |  | (0.001) |  | (0.001) |  | (0.001) |
| Vocational training |  | 0.133** |  | 0.131** |  | 0.119 |  | 0.139 |
|  |  | (0.061) |  | (0.061) |  | (0.085) |  | (0.088) |
| Attended college |  | -0.019 |  | -0.020 |  | -0.015 |  | -0.023 |
|  |  | (0.044) |  | (0.044) |  | (0.063) |  | (0.063) |
| Bachelor's degree |  | 0.007 |  | 0.004 |  | -0.066 |  | 0.078 |
|  |  | (0.042) |  | (0.042) |  | (0.060) |  | (0.058) |
| Graduate degree |  | -0.025 |  | -0.025 |  | -0.053 |  | 0.002 |
|  |  | (0.051) |  | (0.051) |  | (0.073) |  | (0.071) |
| _cons | 0.582*** | 0.414*** | 0.568*** | 0.398*** | 0.568*** | 0.370*** | 0.697*** | 0.546*** |
|  | (0.023) | (0.063) | (0.026) | (0.065) | (0.026) | (0.090) | (0.025) | (0.086) |
| $N$ | 1389 | 1387 | 1389 | 1387 | 705 | 704 | 684 | 683 |
| $R^2$ | 0.015 | 0.027 | 0.016 | 0.029 | 0.008 | 0.031 | 0.001 | 0.017 |

Table 1. Linear probability model regressions of honesty on time pressure. Columns (1) to (4) are based on the full sample, while Columns (5) and (6) are based on the sub-sample of those in the *Own Group Shown* condition and Columns (7) and (8) are based on the sub-sample of those in the *Other Group Shown* condition. Columns (2), (4), (6) and (8) control for gender, age and educational attainment (vocational training, attended college, bachelor's degree and graduate degree – high school or less is the omitted category). Robust standard errors are reported in parentheses. *** $p<0.01$, ** $p<0.05$, * $p<0.1$

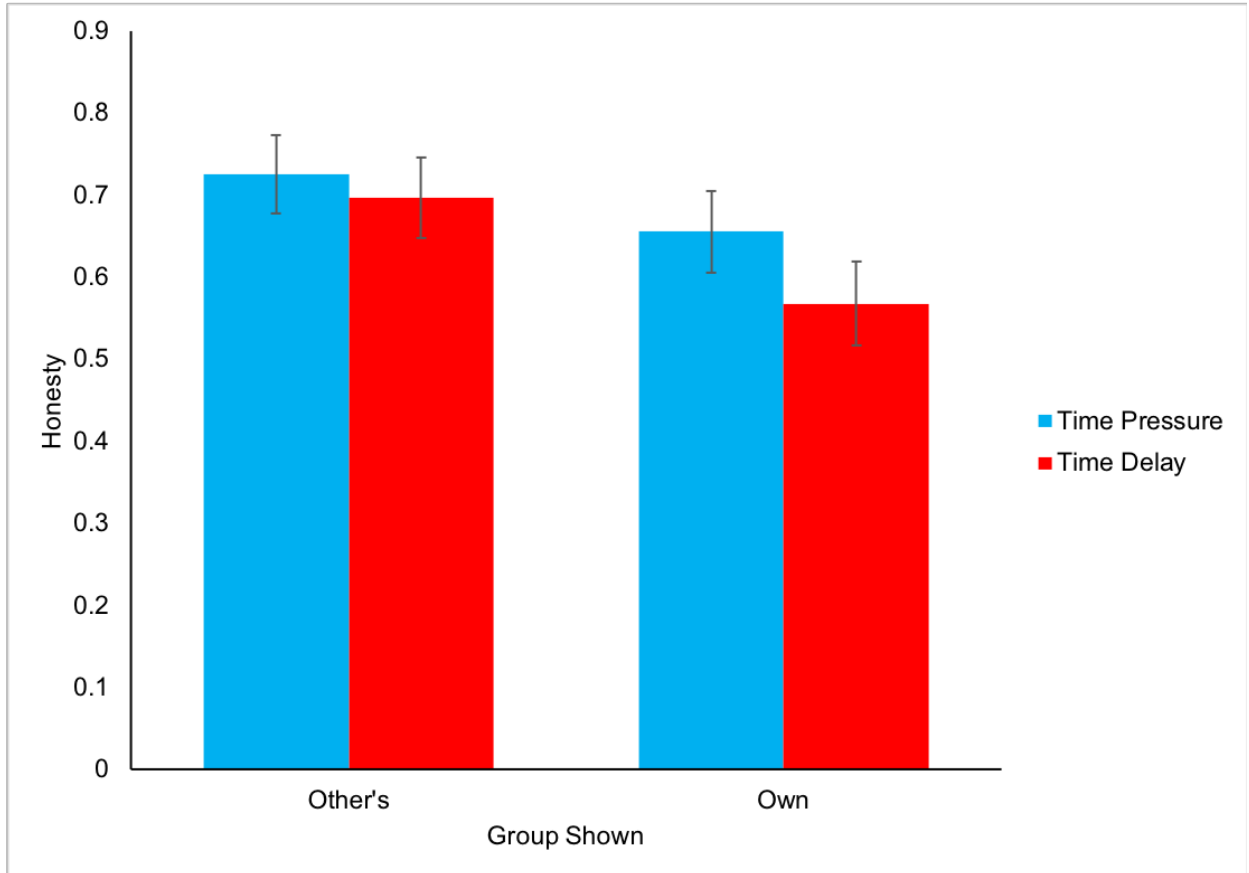

*Figure 1. Average honesty in the time pressure vs time delay condition, separated by whether participants were shown their own group or the group of the participant they were matched with. Error bars represent 95% Confidence Intervals.*

**Discussion**

Whether honesty is intuitive or requires deliberation has attracted considerable attention in the last decade. Several experiments have tried to tackle this question, finding mixed results. Here we have focused on one particular experimental paradigm, introduced by Capraro (2017), which found honesty to be intuitive. The original experiment by Capraro (2017) contains an important confound that can potentially invalidate the interpretation of its results in terms of intuitive honesty. Since participants are shown their group number and have only five seconds to report it either truthfully or not, it might be the case that participants under time pressure are more likely to report whatever number is shown, not because of intuitive honesty, but simply because that number is more readily accessible.

Here, we have improved Capraro's (2017) experimental design in such a way to tell the *Intuitive Honesty Hypothesis* apart from this *Alternative Hypothesis.* Our results are clearly inconsistent with the deflationary hypothesis that pressure just makes people report whatever number is shown, because pressure does not decrease honesty when participants are shown the other's group number. Furthermore, we found no interaction between time pressure and group shown, consistent with the effect of time pressure being similar when one's own group is shown and when the other's group is shown, as predicted by the Intuitive Honesty Hypothesis.

However, the results also raise some questions about the Intuitive Honesty Hypothesis. Although the interaction between time pressure and group number shown is not significant, there seems to be almost no effect of pressure on honesty when participants are shown the matched person's group - which is not what the Intuitive Honesty Hypothesis predicts. Why the time pressure effect seems to decrease when the other group is shown is not entirely clear at this stage of the research. However, we can advance some hypotheses. First of all, we can rule out the hypothesis that our effect is explained entirely by time pressure making people more likely to report the number of the group that is presented to them. If it was the case, average honesty in the *Other Group Shown* condition would be lower than the average honesty in the *Own Group Shown* condition. This prediction is not reflected in the data, which are actually trending in the opposite direction.

This leads us to two potential explanations. One reason for why, in the *Other Group Shown* treatment, the coefficient of time pressure is smaller and not significant could be that this treatment requires a further inferential step for finding out the payoff maximizing strategy, i.e., if the other's group number is 2, then my group number is 1, then to maximize my payoff I have to report that my group number is 2. This additional inferential step requires deliberation and may weaken the role of time pressure. In support of this interpretation, we find that participants take longer to make a decision in the *Time pressure/Other Group Shown* vs the *Time pressure/Own Group Shown* condition (11.24s vs 9.82, rank-sum, $p < .001$). The second potential reason comes from the observation that honesty is significantly higher when subjects are shown the other participant's group number than when they are shown their own. This is a surprising result, that we certainly did not predict, which might suggest that being shown the other person's group makes it salient that lying hurts the other person and thus makes people more honest in general. It is possible that this effect contributes to the lack of an effect of time pressure on honesty in the *Other Group Shown* condition through at least two different paths. Perhaps, people under time delay become even more aware of the fact that lying would hurt another person, and thus become less likely to lie. Another possibility is that there is a ceiling effect: since honesty is higher in the *Other Group Shown* condition, there is less room for statistical changes. Finally, since it is known that M-Turkers tend to be highly experienced (Stewart et al, 2015) and since it is known that experience may decrease effect sizes (Chandler et al, 2015), especially when it comes to cognitive manipulations (Capraro & Cococcioni, 2016; Rand et al, 2014; Rand, 2018), it is possible that our effect is a lower bound of a true effect and that, in reality, time pressure has a positive effect on honesty also in the *Other Group Shown* condition. In any case, we believe that this is an interesting route for future research.

As a side, but still interesting, result, we have also found that men are more likely than women to lie. Gender differences in deception has been largely debated since the early paper by Dreber and Johannesson (2008), which reported that men lie more than women in self-serving situations. Since then, numerous experiments have explored gender differences in (dis)honesty, finding mixed results (Biziou-van-Pol et al, 2015; Cappelen et al, 2013; Childs, 2012; Erat & Gneezy, 2012; Friesen & Gangaradhan, 2012). Two recent meta-analyses shed light on this question: one of experiments using the die-under-cup paradigm (Abeler, Nosenzo & Raymond, in press) and one of experiments using the sender-receiver game (Capraro, 2018), both found that men lie more than women. Our results are in line with these meta-analyses.

In sum, our results support the conclusion of Capraro (2017) that time pressure increases honesty in the deception game studied here. This observation is in line with other work where participants can take actions which help or harm others, namely economic cooperation games (Everett et al 2017; Isler et al 2018; Rand 2016). We conclude by suggesting that the Social Heuristic Hypothesis (Bear & Rand, 2016; Peysakhovich & Rand 2016; Rand et al 2014, 2016), whereby typically advantageous behaviors are internalized as intuitive defaults, may help explain why – and when – honesty is intuitive versus deliberative.

**References**

Abeler, J., Nosenzo, D., & Raymond, C. (in press). Preferences for truth-telling. *Econometrica*.

Ai, C., & Norton, E. C. (2003). Interaction terms in logit and probit models. *Economics Letters*, 80, 123-129.

Arechar, A. A., Gächter, S., & Molleman, L. (2018). Conducting interactive experiments online. *Experimental Economics*, 21, 99-131.

Barcelo, H., & Capraro, V. (2017). The Good, the Bad, and the Angry: An experimental study on the heterogeneity of people's (dis)honest behavior. *Available at SSRN: https://ssrn.com/abstract=3094305*.

Bear, A., & Rand, D. G. (2016). Intuition, deliberation, and the evolution of cooperation. *Proceedings of the National Academy of Sciences*, 113, 936-941.

Biziou-van-Pol, L., Haenen, J., Novaro, A., Occhipinti-Liberman, A., & Capraro, V. (2015). Does telling white lies signal pro-social preferences? *Judgment and Decision Making*, 10, 538-548.

Brañas-Garza, P., Capraro, V., & Rascón-Ramírez, E. (2018). Gender differences in altruism on Mechanical Turk: Expectations and actual behaviour. *Economics Letters*.

Cappelen, A. W., Sørensen, E. Ø., & Tungodden, B. (2013). When do we lie? *Journal of Economic Behavior and Organization*, 93, 258-265.

Capraro, V. (2017). Does the truth come naturally? Time pressure increases honesty in deception games. *Economics Letters*, 158, 54-57.

Capraro, V. (2018). Gender differences in lying in sender-receiver games: A meta-analysis. *Judgment and Decision Making,* 13, 345-355.

Capraro, V., & Cococcioni, G. (2016). Rethinking spontaneous giving: Extreme time pressure and ego-depletion favor self-regarding reactions. *Scientific Reports*, 6, 27219.

Chandler, J., Paolacci, G., Peer, E., Mueller, P., & Ratliff, K. A. (2015). Using nonnaive participants can reduce effect sizes. *Psychological Science*, 26, 1131-1139.

Childs, J. (2012). Gender differences in lying. *Economics Letters*, 114, 147-149.

Dreber, A., & Johannesson, M. (2008). Gender differences in deception. *Economics Letters*, 99, 197-199.

Erat, S., & Gneezy, U. (2012). White lies. *Management Science*, 58, 723-733.

Everett, J. A. C., Ingbretsen, Z., Cushman, F., & Cikara, M. (2017). Deliberation erodes cooperative behavior – Even towards competitive out-groups, even when using a control condition, and even when eliminating selection bias. *Journal of Experimental Social Psychology*, 73, 76-81.

Friesen, L., & Gangadharan, L. (2012). Individual level evidence of dishonesty and the gender effect. *Economics Letters*, 117, 624-626.

Gino, F., Schweitzer, M. E., Mead, N. L., & Ariely, D. (2011). Unable to resist temptation: How self-control depletion promotes unethical behavior. *Organizational Behavior and Human Decision Processes*, 115, 191-203.

Greene, W. (2010). Testing hypotheses about interaction terms in nonlinear models. *Economics Letters*, 107, 291-296.

Gunia, B. C., Wang, L., Huang, L., Wang, J. W., Murnighan, J. K. (2012). Contemplation and conversation: subtle influences on moral decision making. *Academy of Management Journal*, 55, 13-33.

Horton, J. J., Rand, D. G., & Zeckhauser, R. J. (2011). The online laboratory: Conducting experiments in a real labor market. *Experimental Economics*, 14, 399-425.

Isler, O., Maule, J., & Starmer, C. (2018). Is intuition really cooperative? Improved tests support the social heuristics hypothesis. *PLoS ONE*, 13, e0190560.

Lohse, T., Simon, S. A., & Konrad, K. A. (2018). Deception under time pressure: Conscious decision or a problem of awareness? *Journal of Economic Behavior and Organization*, 146, 31-42.

Paolacci, G., Chandler, J., & Ipeirotis, P. G. (2010). Running experiments on Amazon Mechanical Turk. *Judgment and Decision Making*, 5, 411-419.

Peysakhovich, A., & Rand, D. G. (2016). Habits of virtue: Creating norms of cooperation and defection in the laboratory. *Management Science*, 62, 631-647.

Rand, D. G. (2016). Cooperation, fast and slow: Meta-analytic evidence for a theory of social heuristics and self-interested deliberation. *Psychological Science*, 27, 1192-1206.

Rand, D. G. (2018). Non-naïvety may reduce the effect of intuition manipulations. *Nature Human Behaviour*, 2, 602.

Rand, D. G., Brescoll, V. L., Everett, J. A. C., Capraro, V., & Barcelo, H. (2016). Social heuristics and social roles: Intuition favors altruism for women but not for men. *Journal of Experimental Psychology: General*, 145, 389-396.

Rand, D. G., Greene, J. D., & Nowak, M. A. (2012). Spontaneous giving and calculated greed. *Nature*, 489, 427-430.

Rand, D. G., Peysakhovich, A., Kraft-Todd, G. T., Newman, G. E., Wurzbacher, O., Nowak, M. A., & Greene, J. D. (2014). Social heuristics shape intuitive cooperation. *Nature Communications*, 5, 3677.

Shalvi, S., Eldar, O., & Bereby-Meyer, Y. (2012). Honesty requires time (and lack of justification). *Psychological Science*, 23, 1264-1270.

Stewart, N., Ungemach, C., Harris, A. J. L., Bartels, D. M., Newell, B. R., Paolacci, G., & Chandler, J. (2015). The average laboratory samples a population of 7,300 Amazon Mechanical Turk workers. *Judgment and Decision Making*, 10, 479-491.

van't Veer, A. E., Stel, M., & van Beest, I. (2014). Limited capacity to lie: Cognitive load interferes with being dishonest. *Judgment and Decision Making*, 9, 199-206.

# Appendix

## A.1 Testing for Session effects

For logistic reasons we conducted the experiment in two sessions. Here we test whether there are differences between sessions. In table A.1 the regressions include a dummy variable for *Session* (Columns (1) to (8)). In addition, all even numbered Columns control for *Session* interacted with *Time Pressure*, while Columns (3), (4), (7), and (8) also control for *Session* interacted with *Other Group Shown*. Columns (5) to (8) also control for *Time Pressure* interacted with *Other Group Shown*. Table A.1 reveals that all results of the main text are quantitatively similar and that no significant or sizeable difference between the two sessions exist.

|  | Honesty | | | | | | | |
|---|---|---|---|---|---|---|---|---|
|  | (1) | (2) | (3) | (4) | (5) | (6) | (7) | (8) |
| Time Pressure | 0.058** | 0.064* | 0.058** | 0.063* | 0.088** | 0.093** | 0.087** | 0.093** |
|  | (0.025) | (0.035) | (0.025) | (0.035) | (0.037) | (0.044) | (0.037) | (0.044) |
| Other Group Shown | 0.099*** | 0.100*** | 0.087** | 0.087** | 0.129*** | 0.129*** | 0.000 | 0.000 |
|  | (0.025) | (0.025) | (0.035) | (0.035) | (0.036) | (0.036) | 0.116*** | 0.116*** |
| Time Pressure X Other Group Shown |  |  |  |  | -0.060 | -0.060 | -0.060 | -0.060 |
|  |  |  |  |  | (0.050) | (0.051) | (0.051) | (0.051) |
| Session | 0.013 | 0.019 | 0.000 | 0.006 | 0.014 | 0.019 | (0.044) | (0.044) |
|  | (0.025) | (0.036) | (0.037) | (0.045) | (0.025) | (0.036) | (0.037) | (0.045) |
| Time Pressure X Session |  | -0.011 |  | -0.011 |  | -0.011 |  | -0.011 |
|  |  | (0.051) |  | (0.051) |  | (0.051) |  | (0.051) |
| Other Group Shown X Session |  |  | 0.026 | 0.026 |  |  | 0.027 | 0.027 |
|  |  |  | (0.051) | (0.051) |  |  | (0.051) | (0.051) |
| Constant | 0.576*** | 0.573*** | 0.582*** | 0.580*** | 0.561*** | 0.558*** | 0.567*** | 0.565*** |
|  | (0.026) | (0.029) | (0.029) | (0.032) | (0.029) | (0.032) | (0.032) | (0.035) |
| N | 1389 | 1389 | 1389 | 1389 | 1389 | 1389 | 1389 | 1389 |
| $R^2$ | 0.015 | 0.015 | 0.015 | 0.015 | 0.016 | 0.016 | 0.016 | 0.016 |

Table A.1. Linear probability regressions of honesty on time pressure. All columns control for *Other Group Shown* and a dummy variable indicating whether or not the data was collected in Session 2. Columns (5) to (8) control for *Time Pressure* interacted with *Other Group Shown*, Columns (2), (4), (6), and (8) for *Time Pressure* interacted with *Session*, and Columns (3), (4), (7) and (8) for *Other Group Shown* interacted with *Session*. Robust standard errors are reported in parentheses. *** p<0.01, ** p<0.05, * p<0.1

## A.2 Testing for differential effects of *Time pressure* and *Other group shown* depending on demographics

Here we test whether our treatments interact with demographic characteristics. In Column (1) of Table A.2 we interact all demographics (gender, age, educational attainment) with the *Other group shown* condition. We do not find any significant interaction effects for gender, age and educational attainment (only the interaction of *Other Group Shown* and having attained a bachelor's degree is marginally significant). Columns (2) and (3) investigates whether *Time Pressure* and *Other Group Shown* has differential effects depending on the gender. In Column (2) and (3) the regression therefore interacts *Female* with *Time pressure.* In addition, Column (3) contains a three-way interaction of *Female*, *Time Pressure* and *Other Group Shown*. In both specifications we do not find evidence that our treatment variations have a differential effect depending on gender.

|  | Honesty | | |
| --- | --- | --- | --- |
|  | (1) | (2) | (3) |
| Time Pressure | 0.091** | 0.074 | 0.080 |
|  | (0.036) | (0.087) | (0.054) |
| Other Group Shown | 0.176 | 0.130*** | 0.151*** |
|  | (0.125) | (0.036) | (0.053) |
| Time Pressure X Other Group Shown | -0.062 | -0.066 | -0.057 |
|  | (0.051) | (0.050) | (0.075) |
| Female | 0.078** | 0.052 | 0.073 |
|  | (0.037) | (0.036) | (0.053) |
| Female X Other Group Shown | -0.043 |  | -0.042 |
|  | (0.051) |  | (0.072) |
| Female X Time Pressure |  | 0.009 | 0.016 |
|  |  | (0.051) | (0.073) |
| Female X Time Pressure X Other Group Shown |  |  | -0.015 |
|  |  |  | (0.101) |
| Age | 0.003** | 0.002** | 0.002** |
|  | (0.001) | (0.001) | (0.001) |
| Age X Other Group Shown | -0.001 |  |  |
|  | (0.002) |  |  |
| Vocational training | 0.119 | 0.132** | 0.133** |
|  | (0.085) | (0.061) | (0.061) |
| Attended college | -0.015 | -0.020 | -0.019 |
|  | (0.063) | (0.044) | (0.045) |
| Bachelor's degree | -0.066 | 0.004 | 0.005 |
|  | (0.060) | (0.042) | (0.042) |
| Graduate degree | -0.053 | -0.025 | -0.025 |
|  | (0.073) | (0.051) | (0.051) |
| Vocational training X Other Group Shown | 0.020 |  |  |
|  | (0.123) |  |  |
| Attended college X Other Group Shown | -0.008 |  |  |
|  | (0.089) |  |  |
| Bachelor's degree X Other Group Shown | 0.144* |  |  |
|  | (0.083) |  |  |
| Graduate degree X Other Group Shown | 0.055 |  |  |
|  | (0.102) |  |  |
| Constant | 0.370*** | 0.404*** | 0.445*** |
|  | (0.090) | (0.075) | (0.060) |
| $N$ | 1387 | 1387 | 1387 |
| $R^2$ | 0.035 | 0.029 | 0.029 |

Table A.2. Linear probability regressions of honesty on time pressure. Robust standard errors are reported in parentheses.
*** p<0.01, ** p<0.05, * p<0.1

## A.3 Logistic regressions of table 1

Here we report the outcomes of logistic regression of honesty on Time Pressure, Other Group Shown and their interactions. Otherwise the specifications are paralleling those of Table 1 in the main text.

|  | Honesty | | | | | | | |
|---|---|---|---|---|---|---|---|---|
|  | Full sample | | | | Own Group Shown sub-sample | | Other Group Shown sub-sample | |
|  | (1) | (2) | (3) | (4) | (5) | (6) | (7) | (8) |
| Time Pressure | 0.264** | 0.258** | 0.369** | 0.379** | 0.369** | 0.393** | 0.138 | 0.145 |
|  | (0.114) | (0.116) | (0.155) | (0.157) | (0.155) | (0.157) | (0.169) | (0.173) |
| Other Group Shown | 0.448*** | 0.444*** | 0.558*** | 0.570*** | | | | |
|  | (0.115) | (0.116) | (0.159) | (0.161) | | | | |
| Time Pressure X Other Group Shown | | | -0.232 | -0.266 | | | | |
|  | | | (0.229) | (0.232) | | | | |
| Female | | 0.257** | | 0.260** | | 0.337** | | 0.173 |
|  | | (0.116) | | (0.116) | | (0.157) | | (0.173) |
| Age | | 0.011** | | 0.011** | | 0.013** | | 0.009 |
|  | | (0.005) | | (0.005) | | (0.007) | | (0.007) |
| Vocational training | | 0.708** | | 0.704** | | 0.617 | | 0.772 |
|  | | (0.354) | | (0.353) | | (0.461) | | (0.549) |
| Attended college | | -0.086 | | -0.091 | | -0.067 | | -0.106 |
|  | | (0.198) | | (0.199) | | (0.275) | | (0.282) |
| Bachelor's degree | | 0.030 | | 0.020 | | -0.286 | | 0.387 |
|  | | (0.187) | | (0.188) | | (0.259) | | (0.272) |
| Graduate degree | | -0.116 | | -0.116 | | -0.234 | | 0.005 |
|  | | (0.228) | | (0.228) | | (0.316) | | (0.325) |
| _cons | 0.324*** | -0.451 | 0.273** | -0.512* | 0.273** | -0.584 | 0.831*** | 0.092 |
|  | (0.095) | (0.288) | (0.107) | (0.291) | (0.107) | (0.388) | (0.117) | (0.414) |
| N | 1389 | 1387 | 1389 | 1387 | 705 | 704 | 684 | 683 |
| Pseudo $R^2$ | 0.0116 | 0.0220 | 0.0121 | 0.0227 | 0.0060 | 0.0236 | 0.0008 | 0.0143 |

Table A.3. Logistic regressions of honesty on time pressure. Columns (1) to (4) are based on the full sample, while Columns (5) and (6) are based on the sub-sample of those in the *Own Group Shown* condition and Columns (7) and (8) are based on the sub-sample of those in the *Other Group Shown* condition. Columns (2), (4), (6) and (8) control for gender, age and educational attainment (vocational training, attended college, bachelor's degree and graduate degree – high school or less is the omitted category). Robust standard errors are reported in parentheses. *** p<0.01, ** p<0.05, * p<0.1

## A.4 Instructions of the Deception game (common to all participants)

Please read these instructions carefully. Your bonus depends on the decisions you make in the experiment.

You have been matched with another participant. Your bonus and that of the other participant depends only on your choice. The other participant does not play any active role.

There are two groups (group 1, group 2). You will be randomly assigned to one of them. The other participant will be assigned to the other one.

So, if you are assigned to group 1, then the other participant is assigned to group 2. Conversely, if you are assigned to group 2, then the other participant is assigned to group 1.

We will ask you to declare the group you have been assigned to. So you get to choose between two possible options:

Option 1: "I have been assigned to group 1."
Option 2: "I have been assigned to group 2."

If you choose the option corresponding to the number of the actual group you have been assigned to, then you will receive 10c and the other participant will receive 10c.

If you choose the option corresponding to the number of the other group (the one you have not been assigned to), then you will receive 15c and the other participant will receive 5c.

The other participant is real and will really receive the amount corresponding to your choice.

Here are some questions to ascertain that you understand the rules. Remember that you have to answer all of these questions correctly in order to get the completion code. If you fail any of them, the survey will automatically end and you will not get any payment.

What is the choice that maximise YOUR outcome?
(Available answers: Choosing the message corresponding to the number of the actual group you have been assigned to/ Choosing the message corresponding to the number of the other group (the one you have not been assigned to)).

What is the choice that maximise the OTHER PARTICIPANT'S outcome?
(Available answers: Choosing the message corresponding to the number of the actual group you have been assigned to/ Choosing the message corresponding to the number of the other group (the one you have not been assigned to)).

### *Pressure_You_assigned_1 condition*

YOU have been assigned to group 1.

RESPOND WITHIN 5 SECONDS

Which group have YOU been assigned to?

### *Pressure_You_assigned_2 condition*

YOU have been assigned to group 2.

RESPOND WITHIN 5 SECONDS

Which group have YOU been assigned to?

### *Pressure_Other_assigned_1 condition*

The OTHER PARTICIPANT has been assigned to group 1.

RESPOND WITHIN 5 SECONDS

Which group have YOU been assigned to?

*Pressure_Other_assigned_2 condition*

The OTHER PARTICIPANT has been assigned to group 2.

RESPOND WITHIN 5 SECONDS

Which group have YOU been assigned to?

*Delay_You_assigned_1 condition*

YOU have been assigned to group 1.

THINK CAREFULLY FOR AT LEAST 30 SECONDS BEFORE MAKING YOUR CHOICE

Which group have YOU been assigned to?

*Delay_You_assigned_2 condition*

YOU have been assigned to group 2.

THINK CAREFULLY FOR AT LEAST 30 SECONDS BEFORE MAKING YOUR CHOICE

Which group have YOU been assigned to?

*Delay_Other_assigned_1 condition*

The OTHER PARTICIPANT has been assigned to group 1.

THINK CAREFULLY FOR AT LEAST 30 SECONDS BEFORE MAKING YOUR CHOICE

Which group have YOU been assigned to?

*Delay_Other_assigned_2 condition*

The OTHER PARTICIPANT has been assigned to group 2.

THINK CAREFULLY FOR AT LEAST 30 SECONDS BEFORE MAKING YOUR CHOICE

Which group have YOU been assigned to?